# An Empirical Study of the Application of Machine Learning and Keyword Terms Methodologies to Privilege-Document Review Projects in Legal Matters

Peter Gronvall
Data & Technology
Ankura Consulting Group, LLC
Washington, D.C. USA
peter.gronvall@ankura.com

Nathaniel Huber-Fliflet
Data & Technology
Ankura Consulting Group, LLC
Washington, D.C. USA
nathaniel.huber-fliflet@ankura.com

Dr. Jianping Zhang
Data & Technology
Ankura Consulting Group, LLC
Washington, D.C. USA
jianping.zhang@ankura.com

Robert Keeling, Esq.
Complex Commercial Litigation
Sidley Austin LLP
Washington, D.C. USA
rkeeling@sidley.com

Robert Neary
Data & Technology
Ankura Consulting Group, LLC
Washington, D.C. USA
robert.neary@ankura.com

Dr. Haozhen Zhao
Data & Technology
Ankura Consulting Group, LLC
Washington, D.C. USA
haozhen.zhao@ankura.com

*Abstract*— Protecting privileged communications and data from disclosure is paramount for legal teams. Unrestricted legal advice, such as attorney-client communications or litigation strategy. are vital to the legal process and are exempt from disclosure in litigations or regulatory events. To protect this information from being disclosed, companies and outside counsel must review vast amounts of documents to determine those that contain privileged material. This process is extremely costly and time consuming. As data volumes increase, legal counsel employ methods to reduce the number of documents requiring review while balancing the need to ensure the protection of privileged information. Keyword searching is relied upon as a method to target privileged information and reduce document review populations. Keyword searches are effective at casting a wide net but return over inclusive results – most of which do not contain privileged information – and without detailed knowledge of the data, keyword lists cannot be crafted to find all privilege material. Overly-inclusive keyword searching can also be problematic, because even while it drives up costs, it also can cast 'too far of a net' and thus produce unreliable results.

To overcome these weaknesses of keyword searching, legal teams are using a new method to target privileged information called predictive modeling. Predictive modeling can successfully identify privileged material but little research has been published to confirm its effectiveness when compared to keyword searching. This paper summarizes a study of the effectiveness of keyword searching and predictive modeling when applied to real-world data. With this study, this group of collaborators wanted to examine and understand the benefits and weaknesses of both approaches to legal teams with identifying privilege material in document populations.

Keywords: privilege, attorney-client communication, work product doctrine, text classification, predictive coding, technology assisted review, TAR, electronic discovery, ediscovery, e-discovery, privilege review, machine learning

I.  INTRODUCTION

*A. What is Legal Document Review?*

When companies respond to litigation or a request from an enforcement agency (e.g., Department of Justice, Securities Exchange Commission, etc.), they are obligated to produce to the requesting party all non-privileged material relevant to the legal case [1]. To accomplish this, the company's legal teams most often are tasked with gathering, compiling and reviewing large volumes of business documents to determine which documents are relevant to the legal case, and then providing copies of those relevant documents in the form of a document 'production.' Corporations can spend millions or tens of millions of dollars to accomplish this very cumbersome, expensive, and legally required task.

This entire process is referred to in the legal industry as electronic discovery (or 'e-discovery') and more specifically, the 'document review' component of e-discovery. It is an integral part of nearly every sort of litigation or enforcement agency investigation matter in the United States and on an increasing number of countries abroad. Document review requires significant time and resources to meet production schedules established by the legal process; and interestingly, in most instances, production schedules have little regard to the volume of documents at issue.

The costs of document review continue to escalate as the volumes of business data continue to grow. In fact, it can be safely stated that for all practical purposes, data volumes are approaching 'infinite' in size, in the sense that it is essentially impracticable to review each document potentially falling within the scope of an investigation or litigation. And currently, the bulk of e-discovery costs are generated at the document review phase of this process [2].

Traditionally, the document review process begins with collecting data and documents that may contain potentially relevant information. The collected documents are then 'processed' to remove duplicates, to filter by matter-specific criteria (e.g., date ranges, etc.), and finally, to place the documents into an organized, searchable database for attorney review and coding (labeling). When attorneys are reviewing to identify relevant documents, they must also identify which documents contain privileged material. Any documents containing privilege material can and should be withheld from production.

*B. What is Privileged Material?*

Privileged material in today's legal environment are usually emails and electronic documents, either consisting of communications involving lawyers, or documents prepared at the request of lawyers or otherwise in connection to an actual or anticipated legal matter. Privileged materials are also generally 'protected' from disclosure, by the attorney-client privilege and the work product doctrine, both discussed below. Because of these important disclosure protections, it is critical for lawyers to screen for these types of documents to ensure that they are removed from document productions provided to requesting parties.

In the United States of America, there are two important document-disclosure protections: the attorney-client privilege and the work product doctrine.

- The Attorney-Client privilege is a protection from disclosure of confidential communication between an attorney and a client for the purpose of seeking, obtaining, or providing legal assistance [3]. The purpose of this privilege is to ensure free and open communications between attorneys and their clients, without what would be a restraint on those communications, if they were subject to unfettered disclosure.

- Work Product protection is a privilege from the disclosure of material prepared in anticipation of litigation or for trial [4]. The purpose if this privilege is to protect the legal strategy and planning related to ongoing or foreseeable litigation. It also grants attorneys the protected ability to enlist the help of non-attorneys in the preparation of legal representation.

Legal privilege protections provide critically important foundations for the development and creation of sound legal advice and strategy, and they remain an essential part of the U.S. legal system [5]. With origins dating back to the fifteenth century, it has long been established that the construct of legal privilege allows attorneys to be as informed as possible when rendering legal advice [6].

Determining the existence of privilege requires nuance and in-depth calculus when reviewing documents (e.g., relationship between a sender and recipient of communications) and these privileges are not absolute. There are no less than twenty-four actions and situations that can cause the nullification of privilege [6] (referred to in the legal domain as "Waiver") and "few issues arise with greater frequency in civil litigation than whether a document is privileged from compelled disclosure by virtue of the attorney-client privilege." [6]

Legal teams risk nullifying privilege protections if privileged information is missed during the document review process and privilege documents are produced to the requesting or opposing party. The accidental production of privileged communications or work product can be devastating to a legal matter; these documents could provide an opposition party with insight into a company's proposed legal strategy, regulatory decision-making process, or internal investigation findings. A prominent risk of nullification is the application of Subject Matter Waiver, which occurs when privilege material is produced to the opposition. Subject Matter Waiver allows for additional discovery of otherwise protected communications and information which may provide the opposing party with a better understanding of the company's legal strategy. Further, waiver of privilege is not matter specific. When privilege is waived for a specific document or set of documents, it is waived in all future legal proceedings – multiplying the exposure of privileged information.

Historically, the Subject Matter Waiver and risk of future waiver have put significant pressure on legal teams. This pressure drove up the cost of the document review, so much so, that in 2008, the United States Congress amended the Federal Rules of Evidence to limit the scope of Subject Matter Waivers to relieve pressure and reduce the scrutiny legal teams place on privilege material during the document review, thereby reducing document review costs [7].

With data volumes increasing in legal matters, privilege review places a large burden on legal counsel when performing the document review – further increasing cost. Legal teams employ numerous workflows when undertaking privilege document review. Workflows that employ keyword searching to target potentially privileged documents are commonly used to focus the privilege review population and reduce the volume of documents requiring manual review. Typically, counsel develops a list of potentially privileged keyword terms and applies those terms to the relevant document population to identify documents that contain privileged information. These documents are presumed privileged and are reviewed by counsel to confirm their privilege.

### C. Is There a Better Way to Manage the 'Privilege Review' Process?

Yes. In response to the surging volumes – and incumbent costs – of document review, the legal industry is turning to advanced text analytics techniques in search of greater efficiencies and overall accuracy. Most specifically, attorneys and their clients are deploying predictive modeling to identify relevant documents in legal cases. Modeling techniques have been successful at reducing review populations by up to 81.2 percent in legal matters when targeting relevant materials [8]. This reduces the time and cost of attorney review. During a recent regulatory matter, this group of collaborators created a predictive model that targeted relevant content, reduced the volume of review by 64.8 percent, and saved the company more than eight million dollars in document review costs.

While using predictive models to target relevant content has been embraced by the legal community, there is a stigma that predictive modeling cannot reliably identify privileged material. Anecdotally, attorneys maintain a belief that predictive modeling is not precise enough to classify privileged material due to the nuance of specific privilege determinations and the relationship-driven nature of communications. Further, attorneys are reluctant to accept that predictive modeling may yield less than 100% recall – an accepted practice when determining relevance.

As the volume of business data grows [9], the cost to fulfill discovery obligations rises. Legal teams have been successful using traditional relevance culling techniques, such as keyword searching or predictive modeling to reduce the number of documents requiring review and the review costs. However, there is little research, if any, outlining the success legal teams have had using keyword terms to target privileged material. With such little research into the effectiveness of using keyword searching to target privileged material, is it possible that the keyword searching method does not perform as well as the legal community believes?

Additionally, there is little research about the use of predictive models to target privileged information other than Gabriel, et al.'s study – The Challenge and Promise of Predictive Coding for Privilege [10]. Perhaps privilege modeling provides new or more effective methods of targeting privilege information – and that notion deserves a fulsome examination. Further, there are not published studies empirically comparing privilege material targeting methods like keyword searching to predictive modeling. This group of collaborators conducted our research to better understand the strengths and weaknesses of keyword searching and predictive modeling as applied to privilege information. In this paper, we (i) outline the keyword searching and predictive modeling methods; (ii) describe the data set and experiments; and (iii) report our results and findings, highlighting key components that differentiate the two methods.

## II. PRIVILEGE KEYWORD SEARCHING & PREDICTIVE MODELING

### A. Keyword Searching

Keyword searching is a common approach used to target privileged material. A list of keyword terms is created by the legal team and then those terms are searched across the document population to identify documents that contain a term hit. Normally, for instance, these lists are made up of attorney names, law firm names, and other known indicia of legal advice like 'attorney client communication' or 'prepared at the request of counsel'. The performance of keyword searching is linked to the legal team's understanding of the documents and a company's business history [11]. Legal teams with limited knowledge of the documents and business history may develop a term list that is over or under inclusive leading to poor performance.

Keyword searching often yields large numbers of 'false positive' documents to ensure the results are comprehensive. Privilege keyword lists often include "wildcard" syntax to account for variations of word usage. Common terms included in a privilege keyword term list include, "privileged" and "confidential." When combined with a wildcard, the terms become "priv*," and "confid*," which increases the possibility of false positives. For example, documents that contain the word "private" or "confident" will hit on the terms. Privilege keyword lists also include names and email addresses of known attorneys and web domains from known outside counsel. These names,

email addresses, and domains also increase the number of false positives (e.g., individuals with common names, such as Smith, Williams, Brown, or Adams).

Creating effective privilege keyword lists requires that legal teams understand all the legal parties that interact with the company and its employees. Companies that retain multiple outside counsel or have a long history of litigation or investigations by enforcement agencies present a challenge when developing privilege keywords lists. Those companies could require thousands of terms to account for the number of potentially privileged names, words, and legal domains. So, what follows is the challenge of keyword searching for protected communications: without clear insight into all of the key legal counsel and events, the privilege keyword list is by definition incomplete, creating the risk that privileged material could 'survive' the keyword list and make it into the production to the opposing party.

### B. Predictive Modeling

These challenges associated with conventional privilege review remain squarely in the focus of today's innovators and thought leaders in the legal technology industry. A solution proposed by innovators and one that is helping overcome these privilege review challenges is predictive modeling. By applying advanced machine learning techniques to the text of documents, legal teams can automatically classify unreviewed documents into predefined categories of interest (e.g., subject-matter relevance (to the underlying request) or attorney-client privilege/work product). Predictive modeling techniques employ text classification to make a binary choice – relevant or not relevant, privileged or not privileged.

Utilizing supervised learning (e.g., a predictive model based on human reviewed training documents), a predictive model is created by analyzing the textual content of each training document and is then used to rank each document in the corpus with a probability score (0-100) that indicates the likelihood of the presence of privileged material. A higher score indicates greater likelihood that the document contains privileged material and the inverse is true of a lower score.

Predictive models that target privileged documents allow legal teams to prioritize likely privileged documents for review by reviewing high scoring documents first. When utilizing predictive modeling and keyword searching together, counsel has greater insight into the precision of a keyword term before document review begins – something that cannot be determined when using keyword terms alone until after review has concluded. Keyword search terms that have a significant percentage of high scoring document hits may indicate that a keyword term is more precise than a term with a significant percentage of low scoring document hits.

Additionally, predictive privilege models allow legal teams to target documents that have high scores but do not hit on a privilege keyword term. For example, in a recent legal matter this group of collaborators created and deployed a predictive model to identify more than 100 privileged documents that did not hit on a fulsome, well vetted privileged keyword search list. In his example, a law firm name was missed and not included on the privilege keyword term list. While 100 documents may sound like a small number of documents, each missed privileged document presents a degree of risk to a corporation. Ultimately, these documents were withheld from document production thereby reducing the risk of revealing case strategy to the opposing party.

### III. EXPERIMENTAL PROCEDURE

The purpose of the work reported in this paper is to empirically evaluate the effectiveness of keyword searching and predictive modeling for legal privilege review. We conducted experiments using one large data set from a confidential, non-public, real legal matter. This matter contained more than eight million documents coded during a privilege review and was made up of various types, such as email, Microsoft Office documents, PDFs, and other text based types. Attorneys reviewed all documents in our data set and their coding labels provided the ability to evaluate the effectiveness of privilege keyword searching and predictive modeling to target privileged data for this real legal matter.

Table 1: Data Set Statistics

| Total Documents | Privileged Documents | Not Privileged Documents | Richness |
|---|---|---|---|
| 8,715,165 | 536,788 | 8,178,377 | 6.16% |

### A. Keyword Searching Experiments

Our keyword search experiments evaluated the performance of each keyword term from a comprehensive list of keywords developed by attorneys. This list contained more than 5,500 terms and was intended to capture any type of privileged material. The term list consisted of words, such as 'Privileged', 'Legal', 'Attorney Client' as well as terms representing email addresses and law firm names and domains. The keyword term list is confidential and we only report specific term results for terms that are non-confidential.

Our keyword searching experiments evaluated the performance of each keyword term using the precision of each term. The precision of a term was calculated using the attorney review coding (labels) associated with the documents that hit on that keyword term. Each keyword term hit document was coded as Privileged or Not Privileged by the attorneys. The following formula was used to calculate precision:

- **Precision** = a / (a + b)

Where a and b are the number of privileged documents and the number of not privileged documents that hit on the term, respectively. Additionally, the performance of the entire combined list of more than 5,500 terms was calculated. For this analysis, we combined the coding results together for all terms using their Privileged and Not Privileged coding and calculated the overall performance of the term list. For precision, we used the following formula:

- **Precision** = c / (c + d)

Where c and d are the number of privileged documents and the number of not privileged documents that hit all the terms, respectively. Recall for the entire combined list of keyword terms was also computed. The recall of the term list was calculated using the following formula:

- **Recall** = c / e

Where c and e are the number of privileged documents that hit on all the terms and the number of all privileged documents coded by attorneys, respectively.

*B. Privilege Predictive Modeling Experiments*

We conducted two types of predictive modeling experiments. Type One evaluated the performance of a privilege model's ability to effectively target privileged documents and Type Two compared the performance of the privilege model to the performance of the privileged keyword term list. Our performance metrics for this set of experiments were: recall and precision.

*C. Type One Experiments*

Type One experiments were designed to confirm if a predictive model can target privileged content and if so, how well. The predictive model was developed during an active legal matter and its training set design reflects the practical requirements of the matter at that time. The training set was created using two different training populations: (i) training documents identified using a preliminary predictive model trained using privileged and not privileged documents from a similar matter, and (ii) training documents identified using keyword searching.

- Training Population One
  - The Preliminary Predictive Model used to identify Training Population One's documents was generated using 1,338 training documents from a similar but different real legal matter. Table 2 provides a breakdown of the training set documents that were used to build the separate model.
    - **Note**: This preliminary model was created to support the practical requirements of the active legal matter. At that time, we believed using a predictive model to target training documents would create a richer training set.

Table 2: Preliminary Predictive Model Training Set Statistics

| Total Documents | Privileged Documents | Not Privileged Documents | Richness |
|---|---|---|---|
| 1,338 | 669 | 669 | 50% |

  - The model discussed above was applied to our data set and we identified the top 1,000 scoring documents and their family members and provided those to attorneys for review. The training documents identified using the preliminary model are detailed in Table 3.
    - **Note**: A family member is defined as: any attachment to an email.

Table 3: Training Population One's Training Documents

| Total Documents | Top 1,000 Documents | Top 1,000 Documents' Family Members | Privileged Documents | Not Privileged Documents | Richness |
|---|---|---|---|---|---|
| 1,583 | 1,000 | 583 | 1,400 | 183 | 88.4% |

- Training Population Two
  - 4,000 documents that hit on privileged keyword terms were identified and those documents and their family members were provided to attorneys for review.
  - These 4,000 documents plus family members were comprised of two subsets of documents.
    - Subset One: The top 10 terms with the most document hits were used to pull a random sample of 1,000 training documents. The family members of these 1,000 documents were added to Subset One's training population.
    - Subset Two: A random sample of 3,000 documents was generated using the remaining keyword term hits. The family members of these 3,000 documents were added to Subset Two's training population.
  - Training Population Twos' training documents are detailed in Table 4.

Table 4: Training Population Two's Training Documents

| Set Name | Random Sample Size | Random Sample Family Members | Privileged Documents | Not Privileged Documents | Richness |
|---|---|---|---|---|---|
| Subset One | 1,000 | 7,039 | 248 | 7,791 | 3.2% |
| Subset Two | 3,000 | 1,991 | 63 | 4,928 | 1.3% |

The predictive model's combined training population using Training Population One and Training Population Two is detailed in Table 5.

- **Note**: These training documents were identified early in the legal matter's document review and before all documents in the data set were available to sample from. Typically, in a real legal matter, all the documents that require review are not available at the beginning of the matter. Practically, predictive models are created by identifying training data that is available at the time.

Table 5: Predictive Model Training Set Breakdown

| Training Population Name | Total Documents | Privileged Documents | Not Privileged Documents | Richness |
|---|---|---|---|---|
| Training Population One | 1,583 | 1,400 | 183 | 88.4% |
| Training Population Two | 13,030 | 311 | 12,719 | 2.4% |
| **Predictive Model Training Set** | **14,613** | **1,711** | **12,902** | **11.7%** |

The machine learning algorithm we used to create the predictive model was Logistic Regression. One of our prior studies demonstrated that predictive models generated with Logistic Regression perform very well on legal matter documents [8, 12]. Other parameters we used for modeling were bag of words with 1-gram and normalized frequency, and 20,000 tokens were used as features.

A typical supervised learning workflow was used to generate the privilege model. Table 6 details the steps of the workflow.

Table 6: Privilege Model Development Workflow

| 1. Train the model. |
|---|
| 2. Score all documents excluding the training documents used to create the model. |
| 3. Evaluate the results of the model using the attorney review coding (labels) applied to the data set. |

The performance of the model was measured by analyzing the recall of the model at various precision rates including: 50%, 75%, 80%, 90%, and 95%. These are practical precision rates an attorney would want to evaluate during the course of a document review. These rates would help an attorney decide how best to implement a privilege model.

*D. Type Two Experiments*

Type Two experiments were designed to compare the performance of the predictive model to the performance of the keyword search terms. We analyzed their performance in two ways:

- Use a novel and unbiased approach to calculate keyword searching's recall and precision at

various rates and compare its performance to predictive modeling.
- Confirm the number of privileged documents the predictive model identified that did not hit on a term in the keyword term list.

*C. A Novel Approach*

We compared the recalls and precisions of the predictive model and keyword searching at various precision and recall rates that are practically used during document review. To do this, a fair comparison methodology was required. Generally, the performance of keyword searching is calculated at a single precision and recall value pair which is based on treating all the documents hits as a single unit. This is because the precision of the individual keyword terms cannot be calculated until the document review is complete and also because keyword searching does not provide a probability rank for each document hit, unlike a predictive model.

To analyze the recalls and precisions of keyword searching at various precision and recall rates we needed to establish a probability rank for each document hit based on its keyword term hit. This required that we assume an attorney review order or ranking of the keyword document hits. For example,

- Consider a scenario where there are 1,000 keyword document hits that are reviewed in a random order from the first keyword document hit to the last document hit.

- Using the results of this hypothetical review, we could calculate the recall after the first 200 document hits were reviewed as the number of all the privileged documents reviewed up to that point (50 privileged documents) divided by the total number of privileged documents identified during the review (300 total privileged documents).
  - 50 privileged documents / 300 total privileged documents = 16.67% recall

- We could also calculate the precision as the number of all the privileged documents identified after reviewing the first 200 documents (50 privileged documents) divided by the document hits reviewed up to that same point in the review (200 documents).
  - 50 privileged documents / 200 document hits reviewed up to that point = 25.00% precision

Reviewing privilege document hits randomly does not assume some privilege terms are more precise than others. In order to take advantage of the individual precision of keyword terms, to provide a fair comparison between the model and keyword searching, and also provide a novel and practical method to examine the precision of individual terms before review is complete, we created a new approach that uses a "training set" to rank the keyword terms. The approach is outlined below:

1. First, we draw a training set of keyword document hits and have them coded by attorneys with privileged and not privileged labels.
2. Next, we order the keyword search terms in this training set based on the precision of each term, which is defined as the number of privileged documents that hit on a given term divided by the number of all documents that hit on that same term in the training set.
3. Then we order the full list of keyword search terms based on each term's precision in the training set – from high precision to low precision.
   - Search terms that do not have a keyword search hit in the training set are randomly placed at the bottom of the list.

The purpose of this new keyword term ranking approach is to generate a probability rank for each keyword term document hit. This probability rank is similar to a probability score generated by a predictive model and can be used to rank the keyword hits so that precisions and recalls can be calculated at various rates.

Using this novel approach provides a fair comparison between keyword searching and the predictive model since they both use a training set to calculate their precision and recall. Keyword searching's recall and precision for the overall term list was calculated using this method for the comparison in this study. We calculate a given recall and precision value pair by using the privileged and not privileged coding of the documents for a keyword term and all the terms above that term on our list that is now ranked by precision from highest to lowest.

IV. EXPERIMENT RESULTS

In this section, we report and discuss our experiment results using keyword searching and predictive modeling to target privileged documents. We report the various precision and recall rates associated with each approach when applied to our attorney reviewed data set. The process, described in III. EXPERIMENTAL PROCEDURE, was created to evaluate the performance of each text analytics type so that it would be possible to analyze their effectiveness independently and unbiasedly.

A. *Keyword Searching Experiments*

Keyword terms hit on 2,493,846 documents in the data set. The precision of the Keyword Term List was: 20.39% and its recall was: 94.74%. The recall is quite good but the precision is low because there were many term hits that did not contain privileged material.

Evaluating the precision of individual keyword terms provided another opportunity to analyze the results. The performance of a selection of non-confidential terms is detailed in Table 7. These are terms that are commonly used across many legal document reviews to target privileged documents. Note: this data set is confidential and the individual performance of all terms has been excluded from this paper.

Table 7: Performance of Commonly Used Terms

| Term | Privileged | Not Privileged | Precision |
|---|---|---|---|
| Legal | 325,500 | 584,636 | 35.76% |
| Client* | 96,951 | 414,549 | 18.95% |
| Privi* | 189,608 | 315,316 | 37.55% |
| Counsel* | 140,598 | 130,369 | 51.89% |
| Attorney* | 133,914 | 135,004 | 49.80% |

The terms from Table 7, while broad, generally have better precision individually when compared to the precision of the entire Keyword Term List. The only term with less precision is: "Client*". These terms are generally considered very broad and thought off as not precise in the legal community. These results suggest that these terms actually perform well compared to the overall precision of the term list.

To further breakdown the performance of the Keyword Term List, each term was placed into a Precision Band to confirm how many terms belong to each band and the percentage of terms from the term list that belong to each band. The bands create 11 precision ranges that can be used to further aggregate the performance of individual terms. The precision off each term was rounded to a whole number to place the terms into their respective band with the exception of 0 – .99. The 0 – .99 band was rounded to two decimal places. The results of each precision band are detailed in Table 8.

Table 8: Keyword Term List Precision Bands

| Precision Band | Term Count | Band Percentage | Recall |
|---|---|---|---|
| 0% – .99% | 1,717 | 30.95% | 1.09% |
| 1% – 10% | 696 | 12.55% | 12.34% |
| 11% – 20% | 465 | 8.38% | 30.53% |
| 21% – 30% | 337 | 6.08% | 25.85% |
| 31% – 40% | 359 | 6.47% | 81.73% |
| 41% – 50% | 376 | 6.78% | 62.25% |
| 51% – 60% | 334 | 6.02% | 65.76% |
| 61% – 70% | 346 | 6.24% | 28.24% |
| 71% – 80% | 333 | 6.00% | 22.80% |
| 81% – 90% | 269 | 4.85% | 14.89% |
| 91% – 100% | 315 | 5.68% | 9.39% |

**Note**: documents can hit on more than one term. For example, a document could hit on a term in band 91% – 100% and also on a term in band 31% – 40%. This is why the recall across bands does not add up to 94.74%.

Precision band 0% – .99% contained 30.95% of the keyword terms but yielded only 1.09% of all privileged documents. Terms in this band did not effectively identify privileged information. Precision band 31% – 40% contained 6.47% of the keywords terms and yielded 81.73% of all privileged documents. These terms alone were very effective at identifying privileged information.

B. *Predictive Modeling Experiments*

Predictive models developed to target privileged material are typically used to identify attorney review mistakes by revealing documents a model indicates are privileged that an attorney reviewer labeled as Not Privileged. Additionally, privilege models are used to target likely privileged documents that do not hit on a term from the privilege keyword term list. To evaluate the practical implementation of the model, we analyzed the model's performance at high precision levels.

Table 9 details the recall rates of the model at various precision rates that would be used to target privileged documents.

Table 9: Predictive Model Precision Rates

| Precision | Recall |
|---|---|
| 50% | 57.28% |
| 75% | 4.72% |
| 80% | 2.09% |
| 90% | 0.47% |
| 95% | 0.25% |

The model's recall drops significantly from 50% precision to 75% precision. At 95% precision, while the recall is .25%, this is still 21,787 documents that have a high probability of containing privileged material. The model's precision / recall curve is displayed in Figure 1.

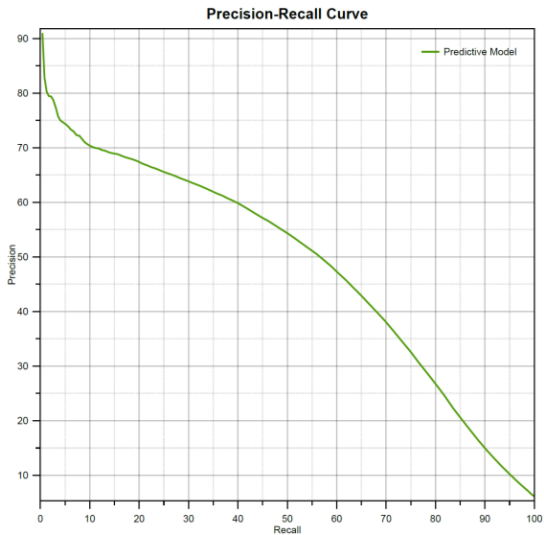

Figure 1: Precision / Recall Curve

C. *Keyword Searching and Predictive Modeling Comparison*

Table 10 and 11 compares the performance of the keyword searching to predictive modeling using the novel approach described in III. EXPERIMENTAL PROCEDURE. Their performance is measured at various precision and recall rates.

Table 10: Recall at Various Precision Various Rates

| Precision | Predictive Modeling Recall | Keyword Searching Recall |
|---|---|---|
| 50.00% | 57.28% | 47.09% |
| 75.00% | 4.72% | 8.05% |
| 80.00% | 2.09% | 3.97% |
| 90.00% | 0.47% | 2.51% |
| 95.00% | 0.25% | 2.51% |

Table 11: Precision at Similar Recall Rates

| Recall | Predictive Modeling Precision | Keyword Searching Precision |
|---|---|---|
| 94.74% |  | 23.29% |
| 94.83% | 10.05% |  |

Overall, keywords yielded greater precision than the predictive model but they did miss key privileged documents that the predictive model did identify. The predictive model identified 11,090 privileged documents that did not contain a privileged keyword. These documents were identified by reviewing any document at a precision of 57.59% and did not hit on a keyword term in order to achieve a recall with the model of 75%. The number of documents requiring review from this population to find the privileged documents was: 89,687. The precision of this sub review was: 12.36%.

V. CONCLUSION

Privileged document review is a complex and risky requirement placed on legal teams. Determining if privilege material is present in document populations requires a nuanced calculus and, as data volumes continue to increase, remains a costly and time-consuming task. Legal teams have traditionally used search terms to target privileged material but there is little research on the effectiveness of that method. Further, the legal community has embraced predictive modeling to target relevant content but has hesitation to use it for targeting privilege content. In this paper, this group of collaborators analyzed the main privilege material targeting methods to provide the legal community with more insight into their performance.

Our experiments empirically compared two privilege material targeting methods in the context of legal document review: keyword searching and predictive modeling. This study examined the performance of keyword searching and predictive modeling using an attorney reviewed data set of 8,715,165 documents, which contained 536,788 privileged documents. We analyzed each privileged material targeting method independently and also compared their performance to each other.

Our study provided many observations with highlights below:

- The keyword searching method's keyword term list was 13.24% more precise than the predictive model. This is surprising given how broad and over inclusive this keyword term list was intended to be. Many privilege term lists are designed to be over inclusive to maximize recall. These results demonstrate that this technique provides higher performing results when compared to predictive modeling.

- Some terms on the keyword term list were expected to have very low precision although they did not. For example, "Counsel*" was 51.89% precise. This term is thought to be very broad in the legal community, yet in this experiment it 2.5 times more precise than the performance of the overall term list.

- The predictive model identified 11,090 privilege documents that did not hit on a privileged keyword term. This represents roughly 2% recall and while that may appear to be a small number, in the context of privilege, this is a huge volume of privileged documents that the attorneys could have missed and possibly produced. Predictive modeling in this case provided a backstop and reduced significant risk.

- The novel keyword search term ranking approach could prove helpful to legal teams looking to reduce the time and cost of a privilege review. Terms identified to have high precision, established after reviewing the "training set," could be used to prioritize the privilege review. Typically, privileged documents go through several rounds of review before the review process completes. Removing one or more rounds of review to confirm privilege could help reduce costs associated with the overall review.

In sum, our study provides insights about the effectiveness of two privileged material targeting methods: keyword searching and predictive modeling. The results of our research suggest that keyword search terms, even broad terms, can achieve high precision and assist with effectively identifying privileged documents. The legal community can be confident in their use of keyword searching to target privileged material going forward. Additionally, our research suggests that privilege models should be used to identify additional privileged content where keyword search terms lists are too narrow and do not cover all parties that generate privileged content. Legal counsel can use this research to create more robust and comprehensive privilege document review strategies.